# A simple INDIUM TIN OXIDE/glass DRA

Vivek Parimi[1], Chia Hao Ku[2], Abhirup Datta[3], Sajal Biring[4], Somaditya Sen*[5,4]

[1]Department of Electrical Engineering, Indian Institute of Technology Indore, Indore, 453552, India
[2]Department of Electrical Engineering, Ming Chi University of Technology, New Taipei, 24301, Taiwan
[3]Discipline of Astronomy, Astrophysics and Space Engineering, Indian Institute of Technology Indore, Indore, 453552, India
[4]Organic Electronics Research Center, Ming Chi University of Technology, New Taipei, 24301, Taiwan
[5]Department of Physics, Indian Institute of Technology Indore, Indore, 453552, India

## ABSTRACT

A novel Dielectric Resonator Antenna, simply made of INDIUM TIN OXIDE coated glass slides placed on a microstrip transmission line, for communication applications is presented. Changes in the bandwidth and gain of the antenna are observed by modifying the dimensions of the INDIUM TIN OXIDE coated glass slides. Changes in gain, directivity and reflection coefficient are observed. A parametric study is conducted on the size of the DRA to understand the effect on bandwidth, reflection coefficient and gain.

*Email:* sens@iiti.ac.in, biring@mail.mcut.edu.tw

## INTRODUCTION

Communication systems operate at 2.45GHz (WLAN), 5.2GHz and 5.8GHz (WiMAX)[1]. WLAN requires low cost and compact antenna with sufficient bandwidth for its application in communication[2]. There are many kinds of antennas, e.g. planar, microstrip, printed mono/dipole, wide-slot, metamaterial, dielectric resonator etc. antennas[3]. Most of these are bulky, difficult to fabricate and cost ineffective. Microstrip patch antennas have narrow bandwidth and larger size[4]. Hence, Dielectric Resonator Antennas (DRA) are being explored. Advantages of DRAs include small size, low cost, high radiation efficiency, reduced conductive losses, and a flexible feed mechanism[5]. Dielectric resonators were reported as filter elements devices in microwave circuits[6]. The smaller size and higher frequency applicability created new interest in DRAs[7]. Today, DRA technology is a promising alternative to conventional antennas for wireless communication applications[8]. Most DRA materials also involve complicated shapes and setup. In this work a simple, cheap and easy to setup DRA antenna, composed of a simple transmission line and an INDIUM TIN OXIDE coated glass slide, is being proposed with considerable gain (GAIN), remarkable bandwidth and significant reflection coefficient ($S_{11}$).

**ANTENNA DESIGN**

The figures above show the geometry of the proposed Dielectric Resonator Antenna (DRA) model. The dimensions of the substrate are length L = 50 mm and width W = 45 mm. The substrate is made of FR-4 and has a dielectric constant of 4.3. The height of the substrate is 1.6mm which is the industrial standard. The gainound plane made of copper has a width W = 45 mm and a length g = 16 mm. The microstrip feed transmission line is made of copper and has dimensions t = 3 mm and length Lt = 42 mm. These dimensions were chosen to create a 50 Ω transmission line. The feeding mechanism of the antenna is a direct microstrip feed. The sample is placed directly over the transmission line at a distance of x = 18 mm from the edge of the board as shown in the figure above. The glass samples used in the experiments are cuboids with square faces having side length ranging from 10 mm to 20 mm. The thickness of all samples is 0.76 mm. The dielectric constant of the glass is 4.82.

The DRAs were placed in two different setups. In one setup the conducting INDIUM TIN OXIDE layer was the top surface. This setup is hereafter referred as T-10, T-15 and T-20 for 10x10mm$^2$, 15x15mm$^2$ and 20x20mm$^2$ slides respectively. In the other setup the bottom surface was the conducting INDIUM TIN OXIDE layer and hence it touched the transmission line. This setup is hereafter referred as B-10, B-15 and B-20 for 10x10mm$^2$, 15x15mm$^2$ and 20x20mm$^2$ slides respectively.

|    | **Parameter**                    | **Value** |
|----|----------------------------------|-----------|
| 1. | Length of substrate (*L*)        | 50 mm     |
| 2. | Width of substrate (*W*)         | 45 mm     |
| 3. | Length of gainound (*g*)         | 16 mm     |
| 4. | Length of transmission line (*Lt*) | 42 mm   |
| 5. | Width of transmission line (*t*) | 3 mm      |
| 6. | Slide position (*d*)             | 18 mm     |

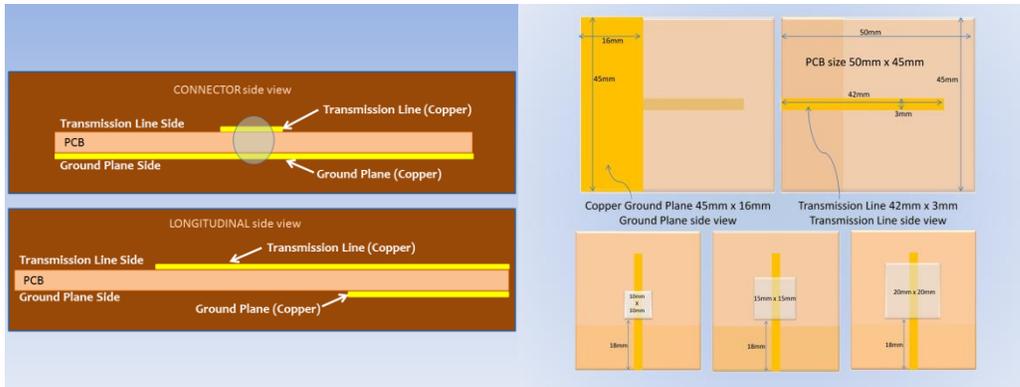

Figure 1: Model and dimensions of the DRA antenna with transmission line, gainound plane and the INDIUM TIN OXIDE coated glass slide arrangement.

**EXPERIMENTAL SETUP**

A 40 GHz Vector Network Analyser (VNA) is used to measure the frequency dependent reflection coefficient ($S_{11}$) of the DRA. The VNA is calibrated for a frequency range of 0.01-12 GHz and is then connected to the antenna using an SMA connector. The scattering parameters are then extracted and used for further analysis.

**ANALYSIS**

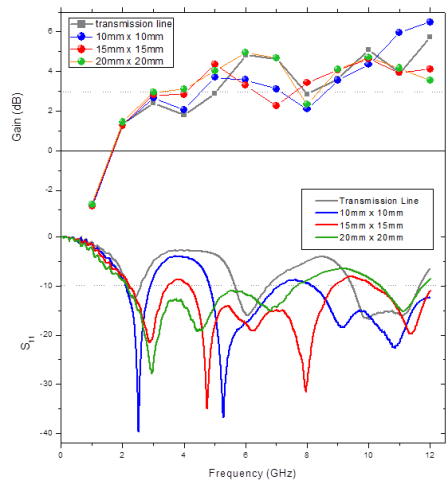

Figure 2: Frequency dependence of gain and reflection coefficient ($S_{11}$) for transmission line and bottom INDIUM TIN OXIDE coated glass slides of the three dimensions.

The transmission line itself has bandwidths of 2.14-2.58 GHz, 5.66-6.7 GHz, and 9.45-11.62 GHz [Figure 2] with considerable gain. Gain for the transmission line is mostly above 2 dB. With the introduction of the INDIUM TIN OXIDE slide, the changes were drastic. Gain values changed nominally and the bandwidth increased considerably.

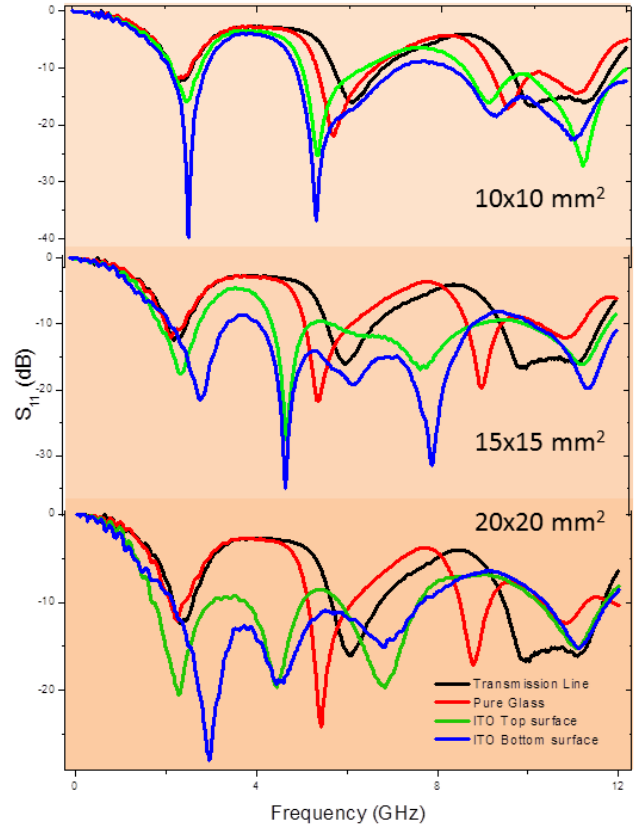

Figure 3: Comparison of reflection coefficient ($S_{11}$) of the transmission line, a pure glass slide, INDIUM TIN OXIDE top coated surface and INDIUM TIN OXIDE bottom coated surface slides of dimension (c)10x10mm$^2$ and (d) 15x15mm$^2$ and (e) 20x20mm$^2$ slide.

For the B-10 slide, the available bandwidths are 2.11-2.91 GHz, 4.84-7.04 GHz, 8.21 GHz and higher. For the B-15 slide, 2.21-3.45 GHz, 4.16-8.87 GHz, 10.16 GHz and higher, while for the B-20 slide, 2.15-7.65 GHz, 10.36-11.78 GHz.

Similarly, the results of T-10, T-15 and T-20 are analogous to B-10, B-15 and B-20. For the T-10 slide, the available bandwidths are 2.16-2.78 GHz, 4.98-6.37 GHz, 8.49-12 GHz. For the T-15 slide, 1.98-2.85 GHz, 4.45-5.31 GHz, 5.75-9.01 GHz and 9.99-11.83 GHz. For the T-20 slide, 1.70-3.01 GHz, 3.76-5 GHz, 5.83-7.578 GHz and 10.22-11.71 GHz.

The responses of the INDIUM TIN OXIDE coated glass slides are much better than both the transmission line. The bandwidth is increased and covers communication bands such as WLAN (2.4-2.5 GHz) and WiMAX (5.15-5.85 GHz).

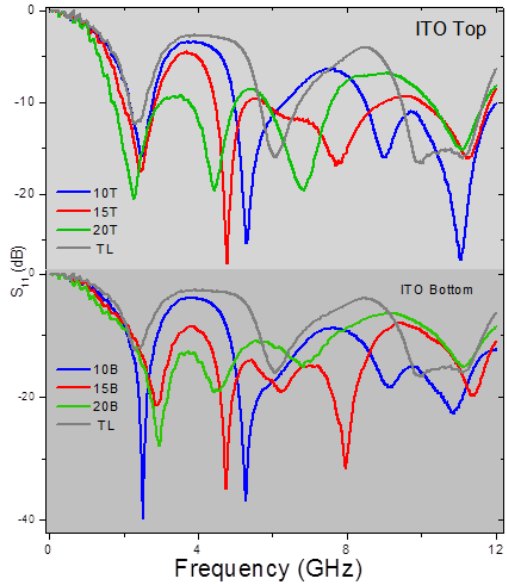

Figure 4: Reflection coefficients ($S_{11}$), measured by VNA, of the INDIUM TIN OXIDE/glass DRA of sizes 10x10mm$^2$, 15x15mm$^2$ and 20x20mm$^2$ with (a) INDIUM TIN OXIDE layer bottom and (b) INDIUM TIN OXIDE layer top

Figures 3 and 4 clearly demonstrate the improvement in the performance of the antenna when INDIUM TIN OXIDE coated glass slides are used. An improvement is seen in both bandwidth and $S_{11}$. Another noteworthy observation is the shift in the frequency of maximum $S_{11}$.

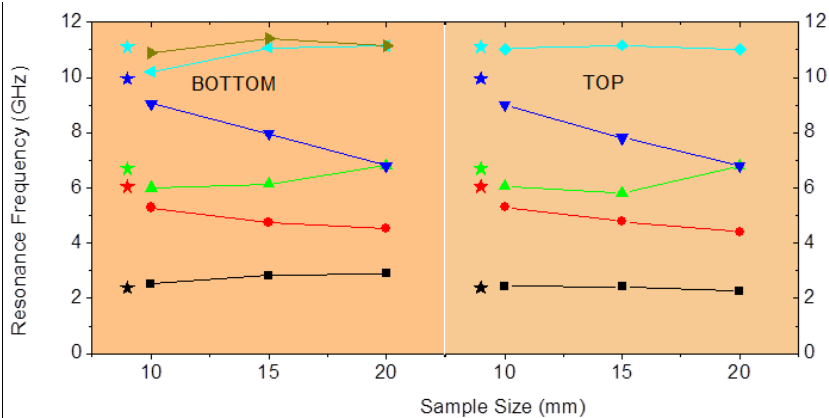

Figure 5: Variation of the resonance frequency (frequency of minima in $S_{11}$) with size of the glass slide. The star denotes the resonance frequency of the transmission line.

The response frequencies for the transmission line itself are at 2.37, 6.04, 6.69, 9.94 and 11.09 GHz. These responses show narrow bandwidths. However, with the INDIUM TIN OXIDE coated glass slides, these responses improve (higher values of $|S_{11}|$) and also shift in position [Figure 5]. Note that the shift is nominal in case of B-10 and T-10 slides. The shifts are more recognized in B-15 and T-15 and most in B-20 and T-20.

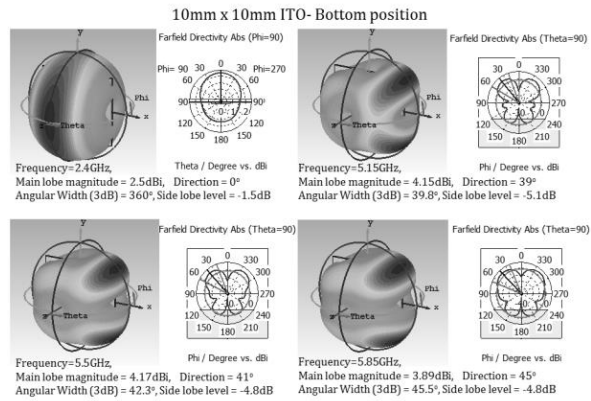

Figure 6: Far field patterns and cuts of B-10 INDIUM TIN OXIDE coated glass slides for 2.4, 5.15, 5.5 and 5.85 GHz frequencies.

In case of B-10 slides, gain of the antenna is considerably higher than 3 dB and is in the range 2.21 to 6.44 dB which makes it highly suitable for communication purposes in the frequency ranges of its bandwidth. It is similar for the T-10 slide. In case of B-15 slides, the gain is slightly lower than that of B-10 slide and ranges from 2.37 to 4.35 dB. The B-15 slide is applicable for slightly lower frequency ranges than B-10 slide. T-15 and B-15 have comparable gain. In case of B-20 slides, the gain of the antenna in the frequency bands of transmission is in the range 2.95-4.7 dB. It has the widest frequency bandwidth and gain is comparable with B-15 and B-10 slides. Similar to the other sizes, T-20 and B-20 are comparable.

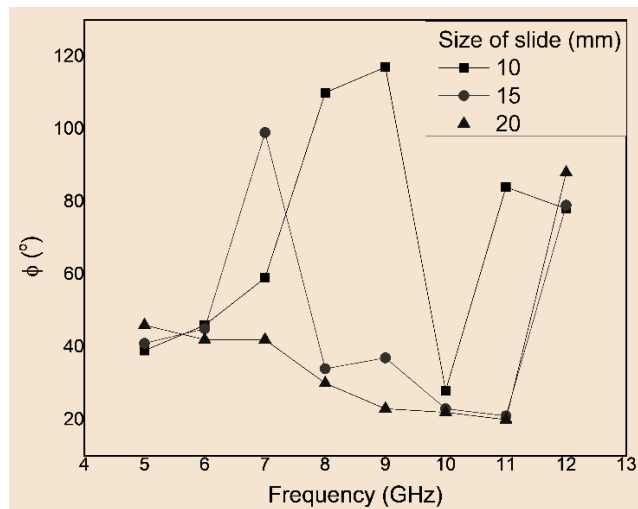

Figure 7: Variation of direction of maximum radiation intensity, $\phi(^o)$ in the xy plane from 5-12 GHz

The radiation direction varies with frequency. At lower frequencies (below 4 GHz), the radiation pattern is like a monopole antenna pattern radiating in the z direction. At higher frequencies, the radiation is due to the higher-order. From 5 GHz to 12 GHz, the radiation is in the xy plane. There are multiple lobes and the direction of maximum radiation (ϕ) varies from $20^o$ to $120^o$ [Figure 7].

Hence, by a simple model comprising of a transmission line and a commercially available INDIUM TIN OXIDE coated glass slide, a competent DRA antenna with high gain and wide band-width can be fabricated in the normal communication ranges. This model is a simple and sure procedure to obtain such novel functionalities.

**CONCLUSION**

An INDIUM TIN OXIDE glass on top of a transmission line can behave as an excellent antenna with gain ranging from 2.21 to 6.44 dB. The antenna can operate in a wide bandwidth from 2.2-12 GHz with considerable $|S_{11}| > 10$ dB. Size dependence of the INDIUM TIN OXIDE coated glass was investigated and it seems that a 20mm x 20mm INDIUM TIN OXIDE coated glass DRA is better than 15mm x 15mm and 10mm x 10mm INDIUM TIN OXIDE coated glass DRAs. It is also observed that if the conducting layer i.e., INDIUM TIN OXIDE layer, is in contact with the transmission line, the response is better with respect to bandwidth and $S_{11}$. A study of the variation of the directivity of the antenna with frequency was conducted by observing the far field patterns and cuts, showing us the variation of direction of radiation at various operating modes of the antenna.